\documentclass[10pt,letterpaper]{article}
\usepackage{Format}
\usepackage[usenames,dvipsnames]{xcolor} 
\usepackage{graphics}
\usepackage{graphicx}
\usepackage{float}
\usepackage{amsfonts}
\usepackage{amsmath}
\usepackage{amssymb}
\usepackage{enumerate}
\usepackage{color}
\usepackage{color,soul} 
\usepackage{multirow} 
\usepackage{rotating} 

\begin{document}

\title{Near-field aperture-probe as a magnetic dipole source and optical magnetic field detector}

\author{Denitza Denkova}
\address{INPAC-Institute for Nanoscale Physics and Chemistry,
Nanoscale Superconductivity and Magnetism and Pulsed Fields Group,
KU Leuven, Celestijnenlaan 200 D, B-3001 Leuven, Belgium}
\email {denitza.denkova@fys.kuleuven.be}
\author{Niels Verellen}
\address{INPAC-Institute for Nanoscale Physics and Chemistry,
Nanoscale Superconductivity and Magnetism and Pulsed Fields Group,
KU Leuven, Celestijnenlaan 200 D, B-3001 Leuven, Belgium}
\address{IMEC, Kapeldreef 75, 3001 Leuven, Belgium}
\email {niels.verellen@fys.kuleuven.be}
\author{Alejandro V. Silhanek}
\address{D\'{e}partement de Physique, Universit\'{e} de Li\`{e}ge, B\^{a}t. B5, All\'{e}e du 6 ao\^{u}t, 17, Sart Tilman, B-4000, Belgium}
\author{Pol Van Dorpe}
\address{IMEC, Kapeldreef 75, 3001 Leuven, Belgium}
\address{Dept. of Physics and Astronomy, KU Leuven, B-3001 Leuven, Belgium}
\author{Victor V. Moshchalkov}
\address{INPAC-Institute for Nanoscale Physics and Chemistry,
Nanoscale Superconductivity and Magnetism and Pulsed Fields Group,
KU Leuven, Celestijnenlaan 200 D, B-3001 Leuven, Belgium}

\begin{abstract}

Scanning near-field field optical microscopy (SNOM) is a technique, which allows sub-wavelength optical imaging of photonic structures. While the electric field components of light can be routinely obtained, imaging of the magnetic components has only recently become of interest. This is so due to the development of artificial materials, which enhance and exploit the typically weak magnetic light-matter interactions to offer extraordinary optical properties. Consequently, both sources and detectors of the magnetic field of light are now required. 

In this paper, assisted by finite-difference time-domain simulations, we suggest that the circular aperture at the apex of a metal coated hollow-pyramid SNOM probe can be approximated by a lateral magnetic dipole source. This validates its use as a detector for the lateral magnetic near-field, as illustrated here for a plasmonic nanobar sample. Verification for a dielectric sample is currently in progress. We experimentally demonstrate the equivalence of the reciprocal configurations when the probe is used as a source (illumination mode) and as a detector (collection mode). The simplification of the probe to a simple magnetic dipole facilitates the simulations and the understanding of the near-field images.
\end{abstract}


\section{Introduction}

The near-fields of photonic nanostructures are characterized by the six components of the electric and magnetic field of light. Thus, the complete understanding of those devices requires measuring of all these six components, the different components having different importance, depending on the specific device and its application \cite {Olmon2010}. At optical frequencies, light interacts with natural materials mainly via its electric field \cite{Landau1960}. Therefore, probing the three electric field components is not only of primary importance, but also much easier than picking up the magnetic field components. One of the most popular ways to do this is by SNOM. In this technique, a probe is positioned in the near-field of the structure and for a specific type of probe and measurement configuration, the electric field distribution can be imaged \cite{M.Schnell2010, Lee2007, BOZHEVOLNYI2000, J.-S.Bouillard2010}. 

However, recently new classes of artificial materials, or so called metamaterials, were developed, for which the magnetic interaction with light becomes as important as the electric one \cite{Liu2008, Rockstuhl2007, Xifre-Perez2013, Smith2000, Soukoulis2011, Dolling2005}. Under these circumstances, both optical magnetic field sources and detectors are now required.

Previously, we have shown that the coupling of a metal covered hollow-pyramid aperture probe to a plasmonic sample results in imaging of the lateral magnetic field distribution of the sample \cite{Denkova2013}. In this paper we suggest that the stand-alone probe can be considered as a tangential optical magnetic dipole source without the need to invoke the coupling to the sample. Additionally, we suggest that the probe can be considered as a tangential optical magnetic field detector. This allows us to substitute the probe in SNOM simulations by an effective magnetic dipole, which significantly decreases the calculation time and memory requirements. Our findings are theoretically justified and validated by numerical simulations of the stand-alone probe and the probe-sample interactions in plasmonic samples. Strong magnetic field contribution to the near-field images obtained on dielectric nanophotonic samples using similar type of probes (matellized aperture optical fiber probes) has recently been reported \cite{Kihm2013, Feber2013}. For the hollow-pyramid probes we use, the verification on other types of samples, such as the above mentioned dielectric samples, is currently in progress.

Recently, also other types of aperture probes, namely, optical fiber aperture probes, were suggested for tangential magnetic field detectors \cite{Kihm2013, Feber2013}. Compared with the optical fiber probes, the hollow-pyramid probes we use \cite{witec, Denkova2013, Denkova2013:2, MicheleCelebrano2009} are more robust (a typical probe under normal operation can last several months) and provide slightly higher resolution topography images of the sample together with the optical magnetic field image. Additionally, the hollow-pyramid probes allow operation in a broad range of wavelengths (500~nm to 1600~nm) with negligible polarization effect on the transmitted light. The drawback is that for metallic samples, a dielectric spacer is recommended on top of the sample to prevent direct ohmic contact between the metal coating of the probe and the sample in contact scanning mode.

First, we will discuss the underlying physics allowing the modelling of the hollow-pyramid probe by a tangential magnetic dipole. Second, we confirm by simulations that the electromagnetic fields of the probe are indeed resembling the ones of a lateral magnetic dipole. Then, we demonstrate that in a simulation describing the real experiment, in which the probe is scanned over a sample (demonstrated for plasmonic bar), similar results are obtained when the probe is substituted by a lateral magnetic dipole. Finally, we experimentally show that the probe can reciprocally be used as a detector for the tangential magnetic field of light.

\begin{figure}[b]
\centering\includegraphics [width=0.75\textwidth] {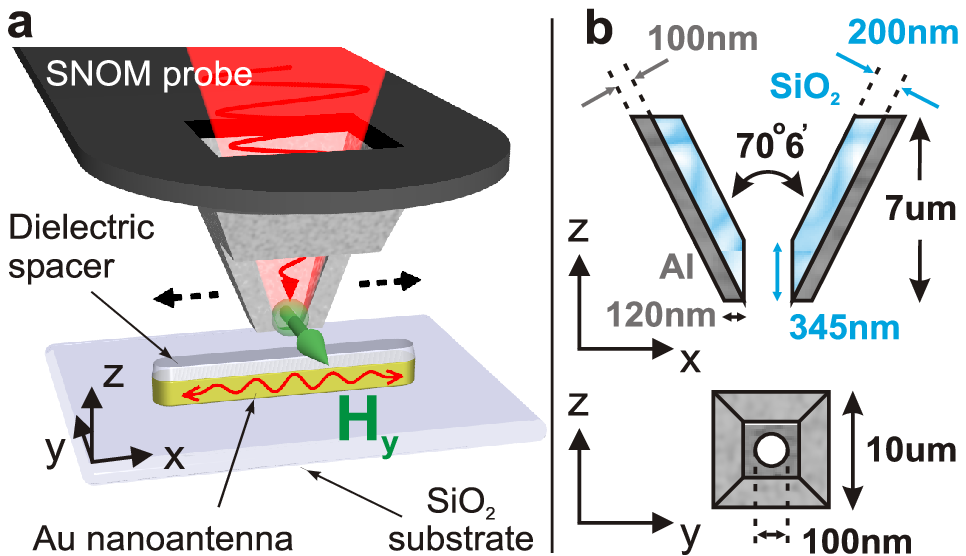}
\caption{(a) The hollow-pyramid probe of a SNOM can be approximated by a magnetic dipole source (green arrow) and, respectively, detector for the tangential magnetic field of light. (b) Structural properties of the pyramid probe, as provided by the probe supplier, are implemented in the FDTD simulations. The figure is not to scale. }
\label{Figure31}
\end{figure}

\section{Results and Discussion}
\subsection{Hollow-pyramid aperture probe as an H$_{tan}$ dipole source: Intuitive physical justification}

Typically, in the near-field measurements, the probe can be approximated by a subwavelength object for which the optical properties are well known in order to facilitate the interpretation of the images and the simulations of the experiments. For instance, dielectric and apertureless metallic scattering SNOM probes, which are used for imaging of the electric field of light are generally modelled as an electric point dipole, with a material and size dependent polarizability \cite{NovotnyBook, Greffet1997, Sun2007, Extarri2009}. A split-ring probe has a strong magnetic dipole moment, normal to the sample surface \cite{M.Burresi2009}. 

Here, we propose that a SiO$_2$ hollow-pyramid metal coated aperture probe can be approximated by an $H_y$ dipole -- a tangential (lateral) \textit{y}-polarized magnetic point dipole, where the \textit{y}-direction is perpendicular to the light polarization direction and to the light propagation direction (Figure \ref{Figure31}).

The specific near-field probe investigated here consists of a hollow SiO$_2$ pyramid, coated with Al, with a subwavelength (100~nm) aperture at its apex (Figure \ref{Figure31} a).  The structure of the probe was modelled following the information provided by the manufacturer (materials used, layer thicknesses, cone angle, and aperture size -- Figure \ref{Figure31} b) \cite{witec}.

To analyse the system, it is useful to simplify the aperture of the hollow-pyramid probe as an aperture in a flat infinitesimally thin perfectly conducting metal screen. The latter is known as a Bethe-Bouwkamp aperture \cite{Bethe1944, Bouwkamp1954} and it has been demonstrated that the transmitted light through such an idealized aperture is similar to the light emitted from a combination of a normal electric dipole and a tangential magnetic dipole \cite{Drezet2002, Kihm2013}. The direction of the magnetic dipole is determined by the polarization of the incident light.

\begin{figure}[b]
\centering\includegraphics [width=0.75\textwidth] {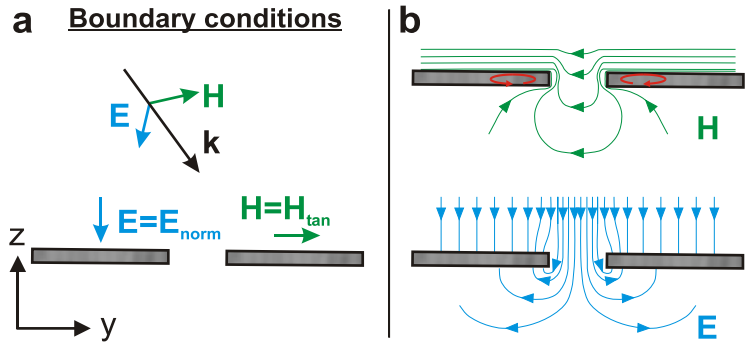}
\caption{The probe can be simplified to a hole in a perfectly conducting metal screen, \textit{i.e.} a Bethe-Bouwkamp aperture. The transmission through such an idealized aperture is similar to the light emitted from a combination of an H$_{tan}$ and E$_{norm}$ dipoles. (a) The boundary conditions at the metal-air interface require the electric field to be perpendicular and the magnetic field to be parallel to the metal. In the vicinity of the hole, this results in the magnetic and electric field line distributions, hand-sketched in (b). The screening currents induced by the magnetic field of the incoming light are depicted in red.}
\label{Figure32}
\end{figure}

One can intuitively picture the situation as follows: The electromagnetic field of the incident light has to satisfy the boundary conditions at the interface between the air and the metal. The general boundary conditions require that at the interface between two materials the tangential electric field E$_{tan}$ and the normal magnetic field H$_{norm}$ have to be continuous. We are considering a perfect metal, which has infinite conductivity, zero skin depth and zero relaxation time, therefore, inside the metal both the magnetic and the electric fields have to be zero. Thus, the fulfilment of the above mentioned boundary condition requires that E$_{tan}$ and H$_{norm}$ components have to be zero also just outside the metal (Figure \ref{Figure32} a).

Due to the fact that we are considering a perfect conductor, which instantaneously reacts to the incoming electromagnetic wave, illumination with any frequency can be assumed as quasi-static. In the quasi-static case Maxwell's equations for the electric and the magnetic fields are decoupled and therefore the response of the metal to those fields can be treated separately. The physical mechanism for screening out the H$_{norm}$ component is the induction of screening currents (red arrows, Figure \ref{Figure32} a) in the metal by the oscillating magnetic field of the incoming light. Respectively, the oscillating electric field is inducing a redistribution of the free charge carriers in the metal to cancel out the E$_{tan}$ component. In Figure \ref{Figure32} b we sketch the electric and magnetic field distributions in the vicinity of the planar aperture to illustrate our intuitive explanation. Rigorous calculations of those fields have been presented elsewhere \cite{Drezet2002, Kihm2013}.  

We have justified that near the air-metal interface the dominant field components are H$_{tan}$ and  E$_{norm}$. Since the hole is small compared to the wavelength of the light, those components are also the main components present in the vicinity of the hole and respectively they are the main components transmitted through the hole, as illustrated in Figure \ref{Figure32} b. Consequently, the radiation transmitted through the hole can be effectively represented by the radiation of an H$_{tan}$ and E$_{norm}$ dipoles.

In our configuration the incident light is mostly perpendicular to the plane of the hole and the metal screen, thus, the E$_{norm}$ component is almost zero. If an E$_{norm}$ component is present, it will induce an effective E$_z$ dipole. Such a dipole does not emit radiation along the z-axis, where our detector is located in the currently used transmission configuration. Therefore, even if it is present, this component will not contribute to the detected signal \cite{Kihm2013}. Kihm \textit {et al.} have shown that in a similar setup, by collecting the polarization resolved light in scattering configuration, the E$_{norm}$ and E$_{tan}$ components can also be measured. In principle, this should also be applicable for our setup. However in the current transmission configuration we would expect our probe to behave only as a tangential magnetic dipole.

In practice, the fine details in the probe structure may slightly modify the result of this idealized simplification and other field components might also get transmitted to some extend \cite{Kihm2013, Feber2013, Drezet2002, Obermuller1995a, Obermuller1995}. Therefore, the assumptions described above need further verification and validation by simulations and experiments.

\subsection{Correspondence between the fields of the hollow-pyramid probe and an H$_{tan}$ dipole: simulations}

In this section we demonstrate that the simulated fields of the hollow-pyramid probe are indeed similar to the fields of an $H_y$ magnetic dipole.

\begin{figure}[b]
\centering\includegraphics [width=0.75\textwidth] {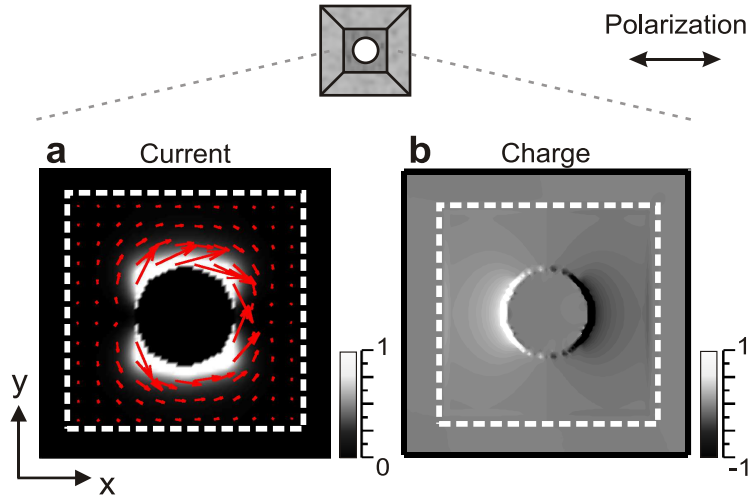}
\caption{Simulations of the real probe show that: (a) The magnetic field of the incident light induces screening currents (red arrows), which in turn induce an effective H$_{tan}$ dipole. The incident light is dominantly perpendicular to the hole of the probe, therefore the contribution from the E$_{norm}$ component is negligible. (b) The specific geometry of the real probe allows a dipolar charge polarization and eventual transmission of the E$_{tan}$ component. The profiles are taken 10~nm above the aperture at $\lambda$~=~1000~nm.}
\label{Figure33}
\end{figure}

Finite-difference time-domain (FDTD) simulations of the probe show that an incident \textit{x}-polarized plane wave generates the above discussed screening currents in the probe, illustrated with the red arrows in Figure \ref{Figure33} a. In turn, those currents induce an effective $H_y$ magnetic dipole, as discussed above for a Bethe-Bouwkamp aperture \cite{Bethe1944}.

\begin{figure*}[b]
\centering\includegraphics [width=1\textwidth] {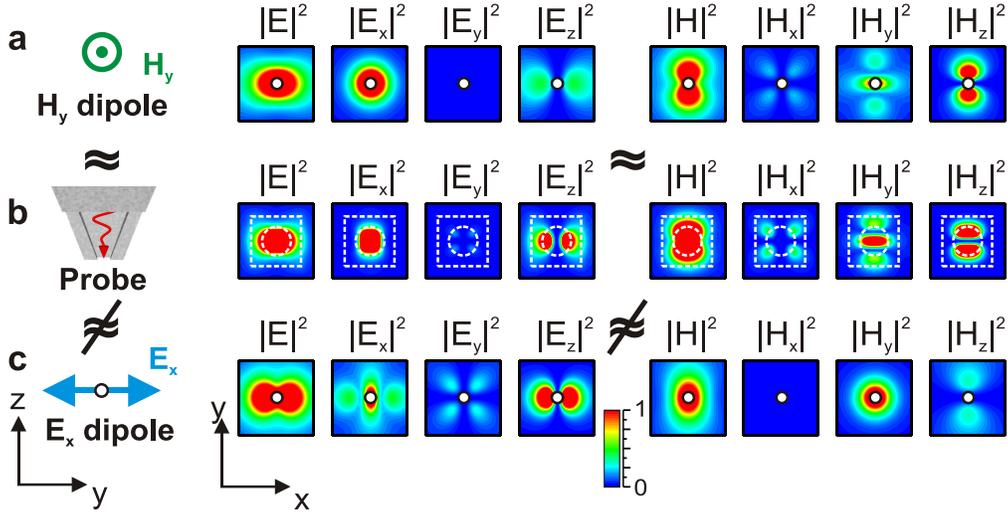}
\caption{The close resemblance of the simulated near-field components of the magnetic dipole (a) and the probe (b) indicates that the probe can be modelled as an H$_{y}$ dipole source. On the contrary, some of the field components of an x-polarized electric dipole (c) differ significantly from the corresponding ones of the probe. Thus, the probe could not be accurately substituted by such a dipole. The probe is outlined with the white dashed lines in (b) and the point dipoles' position is indicated with a white circle in (a) and (c). The x-y field profiles are taken 30~nm below the probe and 130~nm below the dipoles (see horizontal white dash lines in Figure \ref{Figure35}. The simulations are performed at $\lambda$~=~1000~nm.}
\label{Figure34}
\end{figure*}

\begin{figure*}[t]
\centering\includegraphics [width=1\textwidth] {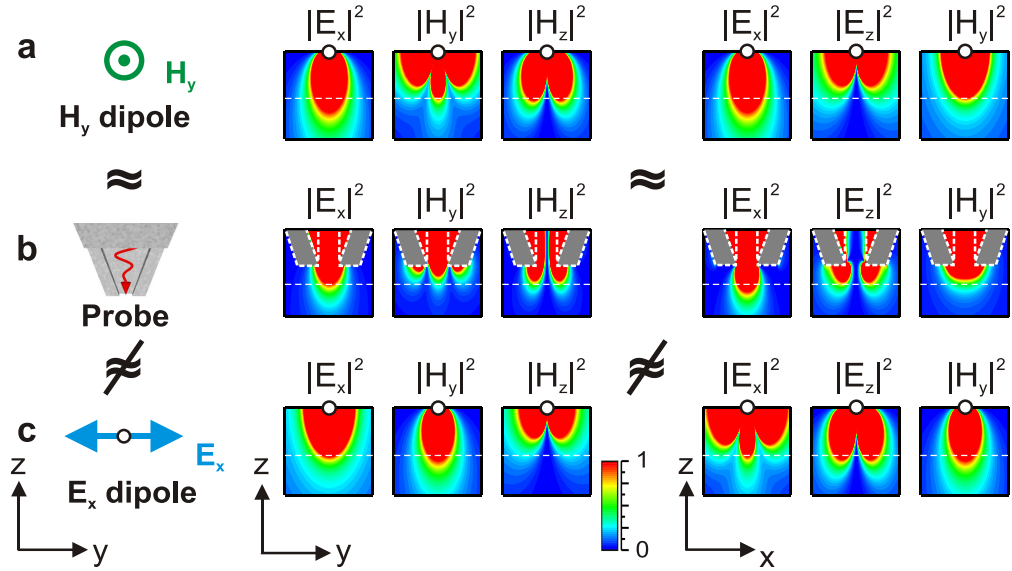}
\caption{The close resemblance of the simulated near-field components of the magnetic dipole (a) and the probe (b) indicates that the probe can be modelled as an H$_{y}$ dipole source. On the contrary, some of the field components of an x-polarized electric dipole (c) differ significantly from the corresponding ones of the probe. Thus, the probe could not be accurately substituted by such a dipole. The x-z and y-z profiles are taken through the middle of the probe/dipoles and only the non-zero field components are plotted. The horizontal white dashed lines indicate the positions at which the x-y cross sections in Figure \ref{Figure34} were taken - 30~nm below the probe aperture and 130~nm below the dipoles. The simulations are performed at $\lambda$~=~1000~nm. }
\label{Figure35}
\end{figure*}

Additionally, the FDTD simulation of the induced charges shows a dipolar charge polarization at the apex of the probe - Figure \ref{Figure33} b. This suggests that the specific probe geometry we use might still allow a certain transmission of the lateral electric field component $E_x$ \cite{Obermuller1995, Obermuller1995a}. Therefore, to check which component is mainly transmitted through the probe and, respectively, by which dipole we can effectively substitute the probe, we compare the simulated probe fields with the simulated fields of those two, $H_y$ and $E_x$, dipoles -- Figure \ref{Figure34} for the \textit{x-y} field profiles and Figure \ref{Figure35} for the \textit{x-z} and \textit{y-z} field profiles. The close resemblance between the near-field of the probe and the $H_y$ dipole supports the idea that the probe can be modelled by an $H_y$ dipole.

The different electric and magnetic near-field components for the probe, as obtained from FDTD simulations, are shown in Figure \ref{Figure34} b. The electric and the magnetic field plots have a common color scale. The probe boundaries are indicated by the dashed white lines. The light incident on the probe is \textit{x}-polarized with wavelength $\lambda$~=~1000~nm. The profiles remain almost unchanged throughout a broad spectral range. The \textit{x-y} cross-sections are taken 30~nm below the aperture and 130~nm below the dipoles (see horizontal white dashed lines in Figure \ref{Figure35}).

Figure \ref{Figure34} a shows the corresponding field components for a \textit{y}-polarized magnetic dipole (H$_y$) and Figure \ref{Figure34} c - for an \textit{x}-polarized electric dipole (E$_x$). The white dot corresponds to the position of the dipole source in the simulation. The strong resemblance of the fields of the probe and the $H_y$ dipole indicates that, in the near-field region, the aperture-probe can be approximated by such an $H_y$ magnetic dipole. In contrast, there is a significant difference in the near field distribution of certain components of the $E_x$ dipole, compared to the probe, particularly for the $H_y$ and the $E_x$ components.

The observation that the near-fields of the probe are similar to those of an $H_y$ dipole (but not an $E_x$ dipole) is also evident in the \textit{x-z} and \textit{y-z} cross-sections - Figure \ref{Figure35}. The cross sections are taken through the aperture center at $\lambda$~=~1000~nm. The horizontal white dashed lines indicate the positions at which the \textit{x-y} cross sections in Figure \ref{Figure34} were taken - 30~nm below the probe aperture and 130~nm below the dipoles. Only the non-zero field components are plotted in the figure.  The simulated profiles in this section are in agreement with results obtained for gold-coated tapered fiber tips \cite{Wei2005}, a planar Bethe aperture \cite{Drezet2004} and a hole in a metal film at THz frequencies \cite{Guestin2009}.

We have now confirmed by simulations our initial assumption that the aperture probe generates strongly resembling electromagnetic fields to those of an $H_y$ dipole. Still, we need to verify that both the probe and the $H_y$ dipole couple similarly to a sample.

\subsection{Scanning of the probe over a sample}

Now we will show that the probe can be substituted by the $H_y$ dipole in a situation mimicking the experiment, \textit{i.e.} where the probe is scanned over a photonic structure. 

The sample we have chosen to demonstrate this consists of a plasmonic antenna gold bar. First, we shortly describe the antenna and the plasmonic effects on it. Then, we will show that the excitation of the plasmon modes in the bar by scanning the probe gives similar results to the excitation of the plasmon modes by scanning an $H_y$ dipole over the bar. Thus, the probe can be modelled as an $H_y$ dipole source also regarding the way it couples to a plasmonic sample. 

The investigated nanoantenna, as shown in Figure \ref{Figure31} a, consists of a 70~nm wide and 50~nm thick gold slab with a 30~nm thick SiO$_2$ capping layer \cite{Denkova2013}. This layer ensures in an experimental setting that the probe is not in direct conductive contact with the sample. The nanoantenna is further supported by a SiO$_2$ substrate. When such a bar is illuminated with light, a surface plasmon wave, \textit{i.e.} a collective surface charge oscillation, is excited. By imposing reflecting boundaries, in this case by confining the plasmon waves in the antenna cavity, standing wave-like Fabry-P\'erot resonances can be formed for certain excitation frequencies \cite{Dorfmuller2009}. These standing wave-like charge oscillations are known as surface plasmon resonances (SPR). The mode index $l$ is used to indicate the number of half plasmon wavelengths $\lambda_p/2$ that fit the antenna cavity at resonance. Consequently, this corresponds to the number of nodes in the plasmon charge density oscillation. 

To initiate this SPR charge oscillation, it suffices to disturb the charge equilibrium. This can be achieved by plane wave illumination or by locally tapping into the electron gas, similar to throwing a stone in a pond. Efficient plasmon excitation, however, is only obtained at specific locations along the standing wave pattern, and intimately depends on the type of excitation source, \textit{i.e.}, its field distribution near the nanoantenna. Since we have shown that the probe fields have a similar distribution as an $H_y$ dipole fields, we can expect that the probe and the $H_y$ dipole will excite the plasmon modes at the same locations along the bar.

\begin{figure*}[t]
\centering\includegraphics [width=1\textwidth] {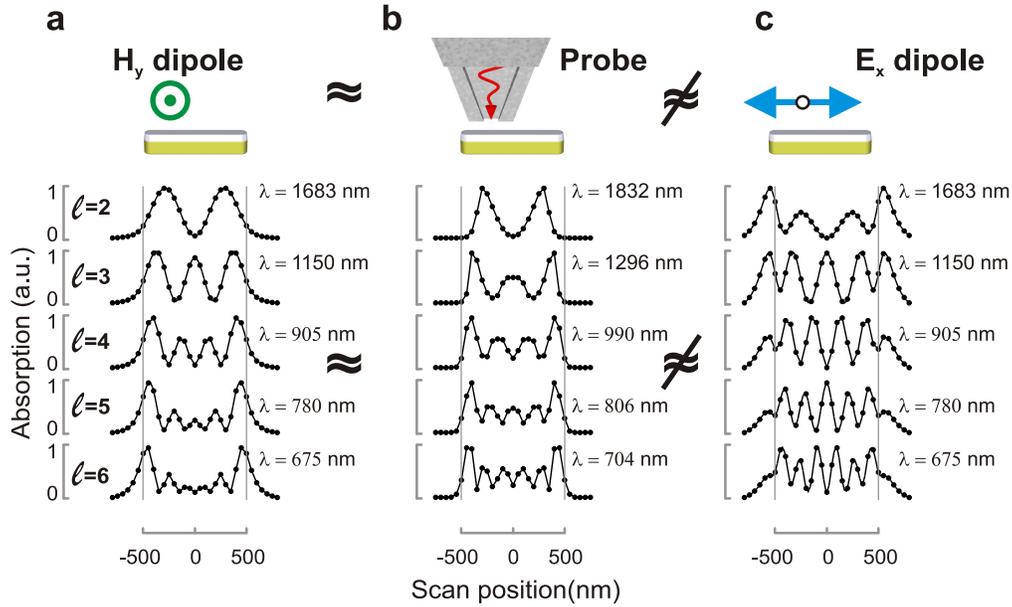}
\caption{The hollow-pyramid circular probe can be modelled by a lateral magnetic point dipole source, as evidenced by the similar absorption profiles obtained when scanning a sample with the probe (b) and an $H_y$ magnetic dipole (a). This is illustrated for the different plasmon resonance modes' absorption profiles in an $L$~=~1~$\mu$m bar. On the contrary, the absorption profiles of the same plasmon modes in the same bar, scanned by an x-polarized electric dipole (c) show additional peaks at the edges of the bar, which are not reproduced in the probe scan profiles (b). The dipoles are scanned at a height 130~nm above the bar. The probe is scanned in contact with the dielectric spacer on top of the bar. The edges of the bar are indicated with the vertical grey lines. Each absorption profile is normalized independently. }
\label{Figure36}
\end{figure*}

As a measure for the excitation efficiency of the plasmon modes we calculated the absorption of the antenna while it is scanned by the probe and the $H_y$ and $E_x$ dipoles described above - Figure \ref{Figure36}. Indeed, the absorption profiles obtained with the probe (b) are similar to the ones obtained with an $H_y$ dipole (a), while very different from the ones with the $E_x$ dipole (c). From top to bottom, the different panels show the absorption profiles for the different order plasmon modes of the bar at the corresponding resonance wavelengths. The \textit{x}-axis indicates the position of the probe along the bar and the \textit{y}-axis gives the corresponding absorption at each of the scanning positions.

In the so-build absorption profiles clear absorption peaks are observed at certain positions - the positions where the corresponding plasmon mode is most efficiently excited. It should be noted that the resonant wavelengths with and without the probe are different, since the proximity of the probe in the near-field of the sample is generating a red-shift in the plasmon resonance wavelengths \cite{Denkova2013}. To facilitate the comparison of the modes, we have normalized each of the modes for each of the sources separately. From the comparison it is clear that the position and shape of the absorption peaks is very similar for the probe and the $H_y$ dipole excitation. The slight differences in the relative intensities and peak shapes is most probably due to the fact that the probe has finite dimensions, while the $H_y$ dipole is a point source. Additionally, weak contributions from other than $H_y$ field components transmitted through the probe, and, effectively inducing other dipole sources, might still be present.

On the contrary, the absorption profiles obtained by an $E_x$ dipole excitation (Figure \ref{Figure36} c) are notably different from the ones obtained by the probe. The $E_x$ dipole clearly fails to properly reproduce the absorption behavior near the edges of the antenna.

Thus, we can safely conclude that the plasmon antenna excitation by the probe can be successfully reproduced by substituting the probe with an $H_y$ magnetic dipole. Such a substitution is in practice very convenient, since the simulations using a single point $H_y$ dipole source are about 30 times faster and 2 times smaller in size than the ones with the probe.

\begin{figure}[t]
\centering\includegraphics [width=0.85\textwidth] {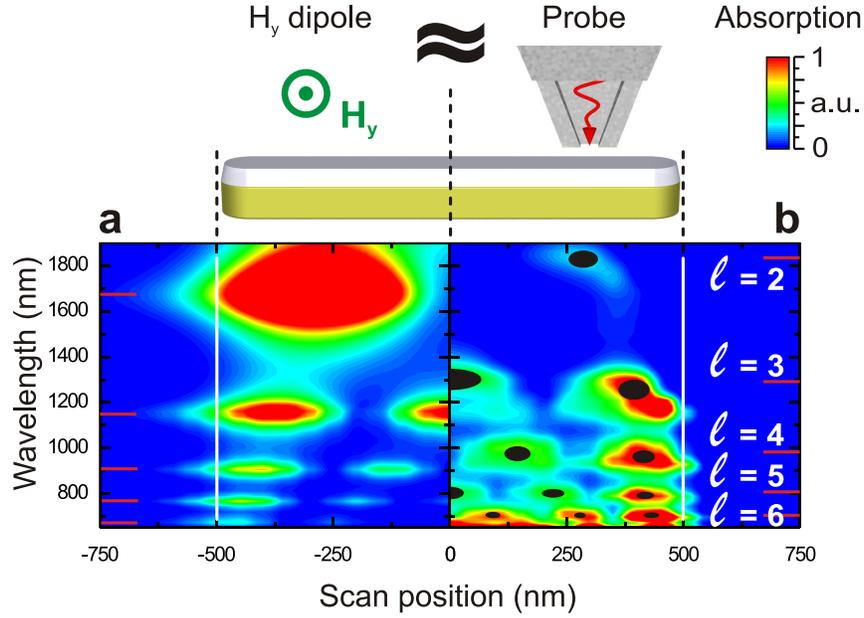}
\caption{The hollow-pyramid circular probe can be modelled by a lateral magnetic point dipole source, as illustrated by the resemblance of the absorption profiles in an $L$~=~1000~nm bar, excited by the probe (b) and the dipole (a) in a broad range of wavelengths. The red lines indicate the plasmon modes' resonant wavelengths for which the profiles in Figure \ref{Figure36} are plotted. The black spots are drawn to highlight the positions of the absorption maxima in the probe scan.}
\label{Figure37}
\end{figure}

The similarity in the plasmon excitation positions for the probe and for the $H_y$ dipole can be observed not only for the absorption profiles at the resonance wavelengths plotted in Figure \ref{Figure36}, but also for other wavelengths. The absorption profiles in the wavelength range between 650~nm and 2000~nm are plotted in Figure \ref{Figure37} for the probe (b) and for the $H_y$ dipole (a). The scans are shown only for the left half of the bar for the dipole and right half of the bar for the probe to facilitate the comparison. As shown in Figure \ref{Figure36}, the other halves have a symmetric profile. The \textit{x}-axis represents the probe position along the bar. The \textit{y}-axis corresponds to the excitation wavelength. The wavelengths at which the scans in Figure \ref{Figure36} are shown are marked with a red line. The color scale goes form blue (low absorption) through green and yellow to red (high absorption).

At the positions along the bar (\textit{x}-axis), where a certain plasmon mode is efficiently excited, the absorption of light with the corresponding wavelength (\textit{y}-axis) is increased - the color scale at those positions is going through green and yellow to red. For clarity, in the right panel the locations at which the plasmon modes are most efficiently excited are highlighted with black ellipses. As expected, at shorter wavelengths more absorption maxima are observed, indicating the excitation of higher order modes in the bar. In this representation it is again visible that the absorption profiles obtained with the probe can be reproduced by an $H_y$ dipole in terms of position and number of absorption maxima. As mentioned before, the near-field proximity of the probe results in a red-shift of the absorption resonances.

The validity of the substitution of the hollow-pyramid probe by a tangential magnetic dipole for scanning other types of samples, for example dielectric samples, is currently in progress. Analogous results have been reported in the literature for similar types of probes -- metallized optical fiber probes \cite{Kihm2013, Feber2013}.

\subsection{Experimental evidence for equivalence of collection and illumination mode SNOM}

We have now shown that the hollow-pyramid SNOM probe can be effectively considered as an $H_y$ dipole source. As discussed in the introduction, not only sources, but also detectors of the magnetic field of light are needed. In the far-field, according to the reciprocity theorem, optical setups with inverted beam paths are equivalent \cite{Born1999}. However, the validity of this theorem in the near-field is not a priori evident and it has to be explicitly demonstrated for the concrete setups \cite{Imura2006, Mendez1997}. Below, we experimentally demonstrate that the inverted beam path schemes of the SNOM setup result in identical experimental images, which means that the hollow-pyramid probe can also be effectively used as an $H_y$ magnetic field detector.  

\begin{figure}[t]
\centering\includegraphics [width=0.6\textwidth] {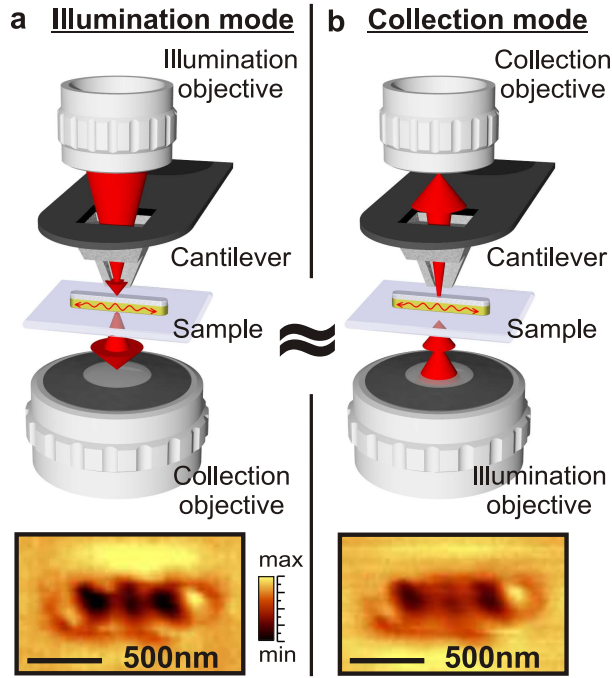}
\caption{Equivalence between illumination (a) and collection (b) mode SNOM. This is confirmed by the identical near-field images obtained in those configurations for the $l$~=~3 plasmon mode of a $L$~=~740~nm bar, illuminated at $\lambda$~=~785~nm.}
\label{Figure38}
\end{figure}

In the illumination mode SNOM, the incoming light is focused through the hollow-pyramid onto the pyramid's apex and the transmitted light is collected in the far field by a collection objective (Figure \ref{Figure38} a). In the reciprocal configuration - collection mode, the nanorod is illuminated by a focused laser beam from the substrate side (at normal incidence to the substrate) and the light picked up through the probe is detected (Figure \ref{Figure38} b). We have carried out this experiment using a gold bar of length $L$~=~740~nm and with illumination wavelength $\lambda$~=~785~nm. The experimental transmission maps present the same characteristic features in both imaging modes - bottom panels in Figure \ref{Figure38} a and b. Three dark regions corresponding to regions of plasmon mode excitation are observed, indicating the excitation of the third order antenna mode.

We have demonstrated previously that the transmission image in illumination mode (Figure \ref{Figure38} a) corresponds to the $H_y$ magnetic field distribution \cite{Denkova2013}. Similarly, Kihm et al. \cite{Kihm2013} have shown for surface plasmon polaritons on a gold film that the experimental images in collection mode (with an optical fiber aperture probe) also visualize the $H_y$ magnetic near-field. In collection mode the image has a slightly reduced contrast, which might be due to slight misalignment -- the alignment procedure is much more challenging from technical point of view. 

Thus, the observed similarity of the near field images in panels a and b indeed confirms the equivalence between illumination and collection modes and the detection of the $H_y$ magnetic field of light by aperture SNOM probes.
The equivalence between the reciprocal configurations of an aperture optical fiber type of SNOM has been reported elsewhere \cite{Imura2006, Mendez1997}.

\section{Conclusions}

In conclusion, we suggest that the circular aperture at the apex of a metal coated hollow-pyramid SNOM probe can be approximated by a lateral magnetic dipole source and, reciprocally, lateral magnetic field detector. This is illustrated by comparing the simulated near-field profiles of the probe with the ones of electric and magnetic dipole sources. We show that the absorption profiles obtained by raster scanning the probe over a metallic sample, in our case a plasmonic bar, are similar to the profiles obtained when substituting the probe by a lateral magnetic dipole. Thus, the probe and the tangential magnetic point dipole source are coupling to the plasmonic sample in a similar way. Currently, this is being verified for other types of samples, namely - dielectric structures \cite{Kihm2013, Feber2013}. This result opens up new possibilities for performing much simpler and faster simulations and gain better understanding of the near-field images. 

We experimentally demonstrate the equivalence between the reciprocal use of the probe as a lateral magnetic field source (illumination mode) and detector (collection mode). This result is practically very useful, because the near-field interactions and images are typically more intuitive to understand when the probe is perceived as a magnetic field detector. However from technical point of view, the setup is much easier to work with and align in the reciprocal illumination mode.

\section*{Acknowledgments}
D.D., N.V., and V.V.M. acknowledge support from the Methusalem funding by the Flemish Government. N.V. acknowledges financial support from the F.W.O.(Flanders). The work of A.V.S. was partially supported by ``Mandat d'Impulsion Scientifique" of the F.R.S.-FNRS. We thank Jos Moonens for his assistance in the e-beam writing.\\


\end{document}